\documentstyle[preprint,aps,epsfig]{revtex}
\draft

\begin{document}

\title{Splitting of Surface Plasmon Frequencies of Metal Particles
in a Nematic Liquid Crystal}
\author{Sung Yong Park and D. Stroud}
\address{Department of Physics,
The Ohio State University, Columbus, Ohio 43210}

\maketitle

\begin{abstract}

We calculate the effective dielectric function for a suspension of
small metallic particles immersed in a nematic liquid crystal
(NLC) host.   For a random suspension of such particles in the
dilute limit, we calculate the effective dielectric tensor exactly
and show that the surface plasmon (SP)resonance of such particles
splits into two resonances, polarized parallel and perpendicular
to the NLC director.  At higher concentrations, we calculate this
splitting using a generalized Maxwell-Garnett approximation, which
can also be applied to a small metal particle coated with NLC. To
confirm the accuracy of the MGA for NLC-coated spheres, we also
use the Discrete Dipole Approximation. The calculated splitting is
comparable to that observed in recent experiments on NLC-coated
small metal particles.

\end{abstract}
\newpage

An important goal in optics is to prepare materials with optical
properties easily controllable by external parameters such as
temperature or electromagnetic fields.  Nematic liquid crystals
(NLC's) are useful materials of this
kind~\cite{lc1,lc2,blinov,lc3}: the optical axis of an NLC, known
as the director, can be rotated by a weak applied electric field
${\cal E}$, thereby altering its transmission and reflection
coefficients.  Several authors have suggested that such control
might be achieved by incorporating NLC's into composite media
which are structured on a scale {\em comparable to the
wavelength}~\cite{busch,zabel,li,kee,kang,shimoda,leonard}.

Recently, there have been new
optical applications of nanoparticles on a scale {\em smaller than
the wavelength}\cite{SPNO}.  Among the applications are
nanolenses~\cite{Li}, plasmonic
nanoantennas~\cite{Genov}, and light energy transfer through linear
chains of nanoparticles~\cite{quinten,maier,maier1}.   Thus,
it would be of great interest if the use of NLC's to control
light propagation could be extended to systems with structure on this scale.
Indeed, a recent experiment has shown that the optical properties of
NLC-coated metal nanoparticles can be controlled by a dc electric
field~\cite{mueller}.

In this Letter, as a first step towards a theoretical treatment of
such systems, we present an exact calculation of the optical
properties of a dilute suspension of metallic nanoparticles in an NLC host.
We find that the usual surface plasmon (SP) resonance of the metal
nanoparticles splits into two resonances, polarized parallel and
perpendicular to the NLC director. In addition, we develop a
Maxwell-Garnett approximation (MGA) which is suitable either for
higher concentrations of particles in an NLC host or for a single
nanoparticle coated with an NLC.  We also use the
discrete dipole approximation (DDA) to confirm the accuracy of the MGA
for NLC-coated spheres.  For typical NLC parameters, we find
that the calculated SP splitting for coated spheres is large enough to be
experimentally observable in both absorption and transmission,
and, indeed, is of a magnitude comparable to that found
in experiments~\cite{mueller}.

First, we consider the optical
properties of a dilute suspension of spherical metallic
nanoparticles in an NLC host, as illustrated schematically
in Fig. 1 (a). The metal
particles are assumed to have a complex but isotropic dielectric
function $\epsilon_m(\omega)$, which we take simply to be
of the Drude form,
\begin{equation}
\epsilon_m(\omega) = 1 - \omega_p^2/[\omega(\omega+ i/\tau)],
\label{eq:drude}
\end{equation}
where $\omega_p$ is the plasma frequency and $\tau$ is a
characteristic relaxation time. The NLC host is taken to be
a uniaxial dielectric with dielectric tensor
$\tilde{\epsilon}_{NLC}$ having principal components
$\epsilon_\perp \equiv n_\perp^2$, $\epsilon_\perp$, and
$\epsilon_\| \equiv n_\|^2$, assumed real and frequency-independent.
If the metal particle radius is small compared to
either its skin depth or the wavelength $\lambda$, the composite
optical properties are well described by a complex effective
dielectric tensor, $\tilde{\epsilon}_e(\omega)$.
We also assume that the NLC is homogeneous, with principal axes
which point in the same directions at every point in the space it
occupies.  (The limitations of this assumption are discussed
briefly below.)  With these approximations,
$\tilde{\epsilon}_e(\omega)$ is also a tensor with principal
components $\epsilon_{e,\perp}(\omega)$,
$\epsilon_{e,\perp}(\omega)$, and $\epsilon_{e,\|}(\omega)$.

We calculate $\tilde{\epsilon}_e(\omega)$ in the
limit that the volume fraction of inclusions $p \ll 1$.
In actuality, $\epsilon_e$ for this
geometry has been calculated in Ref.\ \cite{stroud76} for the
formally analogous case of conducting particle in a host with an
anisotropic d. c. conductivity. The result for the dielectric
function is
\begin{equation}
\tilde{\epsilon}_{e}(\omega) = \tilde{\epsilon}_1 + p[\tilde{1} -
\delta\tilde{\epsilon}\tilde{\Gamma}]^{-1}\delta\tilde{\epsilon},
\label{eq:dilute}
\end{equation}
where
$\delta\tilde{\epsilon} = \tilde{\epsilon_2} - \tilde{\epsilon_1}$
is the difference
between the diagonal dielectric tensors $\epsilon_2$ and
$\epsilon_1$ of the inclusion and the host,
and $\tilde{\Gamma}$ is a
3 $\times$ 3 ``depolarization tensor'' which is diagonal in the
same frame of reference as $\tilde{\epsilon_1}$.
In the case of Fig.\ 1(a), $\tilde{\epsilon}_1$ corresponds to
$\tilde{\epsilon}_{NLC}$
and $\tilde{\epsilon}_2$ to $\epsilon_m\tilde{1}$, where
$\tilde{1}$ is a 3 $\times$ 3 unit matrix. Equation (\ref{eq:dilute}) is a
matrix equation for $\tilde{\epsilon}_e$ but because both
$\tilde{\Gamma}$ and $\delta\tilde{\epsilon}$ are simultaneously
diagonalizable, it is also diagonal and reduces to three separate
algebraic equations for the three components of
$\tilde{\epsilon}_e$. If the metallic inclusions are spherical
particles, then $\tilde{\Gamma}$ has principal components
$\Gamma_\perp$, $\Gamma_\perp$, and $\Gamma_\|$, where
\begin{equation}
\Gamma_\| = -(1 - \sqrt{1 - g}\sin^{-1}\sqrt{g}/\sqrt{g})
/(\epsilon_{\|}g), \label{eq:gampar}
\end{equation}
and
\begin{equation}
\Gamma_\perp = -\frac{1}{2}(\Gamma_\| +
\sin^{-1}\sqrt{g}/\sqrt{g\epsilon_\|\epsilon_\perp}),
\label{eq:gamperp}
\end{equation}
where $g = 1 - \epsilon_\perp/\epsilon_\|$.

The positions $\omega_\alpha$ ($\alpha = \|$ or $\perp$) of the
two peaks can be obtained analytically from the equation
\begin{equation}
1 -\Gamma_\alpha\delta\epsilon_\alpha = 0, \label{eq:det_omega}
\end{equation}
where $\alpha = \|$ or $\perp$.  For a Drude dielectric function
in the limit $\omega_p\tau\rightarrow \infty$, this equation
reduces to
\begin{equation}
\omega_{0,\alpha}^2/\omega_p^2 = -\Gamma_\alpha/
[1 -(1-\epsilon_\alpha)\Gamma_\alpha],
\end{equation}
where $\Gamma_\alpha$ is given by Eqs.\ (\ref{eq:gampar}) and
(\ref{eq:gamperp}).  To present illustrative results,
we assume that the
host is the NLC known as E7, which has principal indices of
refraction $n_\perp = \sqrt{\epsilon_\perp} = 1.52$, $n_\| =
\sqrt{\epsilon_\|} = 1.75$, as used in the experiments in
Ref.~\cite{mueller}. For these $n_\perp$ and $n_\|$,
$\omega_{0,\alpha}/\omega_p = 0.4068$ and $0.4022$. Here the difference
is a small but observable amount, and is robust against small changes of
either index of refraction. It is of interest to compare the
calculated splitting with that observed in the experiments. For
$p = 0.01$, our calculated splitting between the $\|$ and $\perp$ SP
frequencies is about 1\%.  By comparison, that observed in Ref.\
\cite{mueller} appears to be around 3.4\%.

At higher concentrations, $\tilde{\epsilon}_e$ can only be
computed approximately, e.\ g., using the Maxwell-Garnett
approximation (MGA). The MGA is believed to be best suited to a
geometry in which the inclusions are preferentially surrounded by
the host (NLC, in this case).   If the director $\hat{n}$ of the
NLC is independent of position, then the MGA takes the form
$\tilde{\epsilon}_{MGA} = \epsilon_2 +
p\left[\tilde{1} - (1-p)\tilde{\Gamma}\delta\tilde{\epsilon}
\right]^{-1}\delta\tilde{\epsilon}$.

Next, we consider the optical properties of a dilute suspension
of {\em coated} particles.  The coating is assumed to consist
of a thin layer of NLC.   The geometry in this case is shown
schematically in Fig.\ 1(b).  We may calculate the effective
dielectric function of this suspension in two stages.
First, we calculate the dielectric function
$\tilde{\epsilon}_{coat}$ of the coated
particle using the MGA, and assuming that the director $\hat{n}$
in the coating is uniformly oriented.  Thus, we take
\begin{equation}
\tilde{\epsilon}_{coat} = \tilde{\epsilon}_{NLC} + p'\left[\tilde{1} -
(1-p')\tilde{\Gamma}\delta\tilde{\epsilon}\right]^{-1}
\delta\tilde{\epsilon},
\label{eq:epscoat}
\end{equation}
where $s$ is the thickness of the coating, $a$ is the
radius of the included metallic particle,
$p' = a^3/(a+s)^3$ is the volume fraction of metal in
the coated particle, and
$\delta\tilde{\epsilon}=\epsilon_m\tilde{1}-\tilde{\epsilon}_{NLC}$.
In the second stage, we assume that a volume fraction
$p \ll 1$ of coated particles is embedded in an air
host, and calculate the effective dielectric tensor $\epsilon_e$ of this suspension
using eq.\ (\ref{eq:dilute}). To carry out this latter calculation,
we take $\tilde{\epsilon}_1 = \tilde{1}$,
$\delta\tilde{\epsilon}=\tilde{\epsilon}_{coat}-\tilde{1}$, and
use $\tilde{\Gamma} = -\tilde{1}/3$, the depolarization tensor for
a spherical particle in an isotropic host of unit dielectric constant.
The resulting $\tilde{\epsilon}_e$ is diagonal with
nonzero components
\begin{equation}
\epsilon_{e,\mu} = 1 + p t_\mu,
\end{equation}
where
\begin{equation}
t_\mu = \frac{\epsilon_{coat,\mu} - 1} {1 +
(1/3)(\epsilon_{coat,\mu}-1)},
\end{equation}
and the subscript $\mu = \|$ or $\perp$ denotes a component in the
direction parallel or perpendicular to $\hat{n}$.  By analogy with
Eq.\ (\ref{eq:det_omega}), we can determine the positions
$\omega_\mu$ of the two peaks analytically from
the equation
\begin{equation}
1 + (1/3)(\epsilon_{coat,\mu}-1)= 0, \label{eq:mga1}
\end{equation}
where $\mu = \|$ or $\perp$.

We have used Eq.\ (\ref{eq:mga1}) to obtain the SP resonance
frequencies of a dilute suspension of NLC-coated metal particles.
We consider two different angles between the director and the
polarization of incident light, namely $0^o$ and $90^o$. We use
the Drude dielectric function of Eq.\ (\ref{eq:drude}) for
$\epsilon_m(\omega)$, in the limit $\omega_p\tau \rightarrow
\infty$, and the dielectric tensor of E7 for that of the coating.
The resulting frequencies of the SP resonances are shown in Fig.\
2 as full and dashed lines.

To see how well this MGA approach works for a dilute suspension of
coated metal particles, we have compared the results to DDA
calculations of the normalized peak position of scattered
amplitude~\cite{dda}. In the DDA, the scattering object is
represented as a collection of point dipoles on a simple cubic
grid of $n^3$ points, with $n$ as large as 120; we increase $n$
until the peak position of scattered amplitude converges. The
magnitudes of the point dipoles are chosen to reproduce the
scattering properties of the original medium~\cite{polar}.We use
the Drude dielectric function of Eq.\ (\ref{eq:drude}) for
$\epsilon_m(\omega)$, with $\omega_p\tau =7.7$, and the dielectric
tensor of E7 for that of the coating. The peak positions,
calculated in the DDA and interpreted as SP frequencies, are shown
in Fig.\ 2 as open squares and diamonds. These calculations show
that, for $p^\prime \gtrsim 0.5$, the MGA and DDA results are very
close. For $a=30nm$ this result means that for $s \lesssim 8 nm$,
the MGA result is reasonable.  Thus, the Maxwell-Garnett approach
is fairly accurate for the optical response of nanoparticles with
a thin NLC coating, when the director is uniform in the coating.

Finally, we discuss the limitation of our present calculations. In
a plausible NLC layer thickness range, the calculated splitting in
Fig.\ 2 between the SP frequencies for light polarized parallel
and perpendicular to $\hat{n}$ is 1$\sim $2\%, roughly comparable
to experiment~\cite{mueller}. However, to make a detailed
comparison to experiment, we would need to relax some of the
assumptions made in the present calculations; the assumption that
the NLC is homogeneous, near the surface of a colloidal particle,
may be very difficult to achieve experimentally, since in real
experimental situations, the NLC director is substantially
perturbed near the surface of a colloidal particle. The
characteristics of the surface of the particle, which are referred
to as anchoring conditions, can yield various nontrivial
configurations of the nematic liquid crystal~\cite{lubensky};
these anchoring conditions will vary with the experimental
preparation~\cite{experimental}. Thus, a more quantitatively
accurate theory of the optical properties of suspensions should
include these effects, though the present approach should provide
a qualitatively reasonable first step. Calculations which take
into account these complications will be presented
elsewhere\cite{park1}.

This work has been supported by NSF grant DMR 01-04987.  We would
like to thank Dr. S. Maier for useful conversations. Numerical
calculations in this paper were carried out using the facilities
of the Ohio Supercomputer Center.

\newpage
\begin{center}
\large FIGURE CAPTIONS
\end{center}

FIGURE 1. Schematic diagram of the geometries we consider here. (a)
metal nanoparticle embedded in an NLC host. (b) NLC coated
metal nanoparticle in air host.

FIGURE 2. Calculated surface plasmon frequencies
$\omega_{0,\|}(p')/\omega_0$ and $\omega_{0,\perp}(p')/\omega_0$,
corresponding to incident light polarized parallel and
perpendicular to the NLC director, for metal nanoparticles coated
with NLC and embedded in an air host, plotted as a function of
volume fraction $p^\prime$ of the coated particle which is
composed of metal.  The coating is assumed to be the NLC E7, which
has indices of refraction $n_{\|} = 1.75$, $n_\perp = 1.52$.
The metal particles are taken to have the Drude dielectric
function, and $\omega_0 \equiv \omega_0(p^\prime=1)$. Full and
dashed lines: MGA; squares and circles: DDA.

\begin{figure}
\begin{center}
\epsfxsize=11cm \epsfysize=8cm \epsfbox{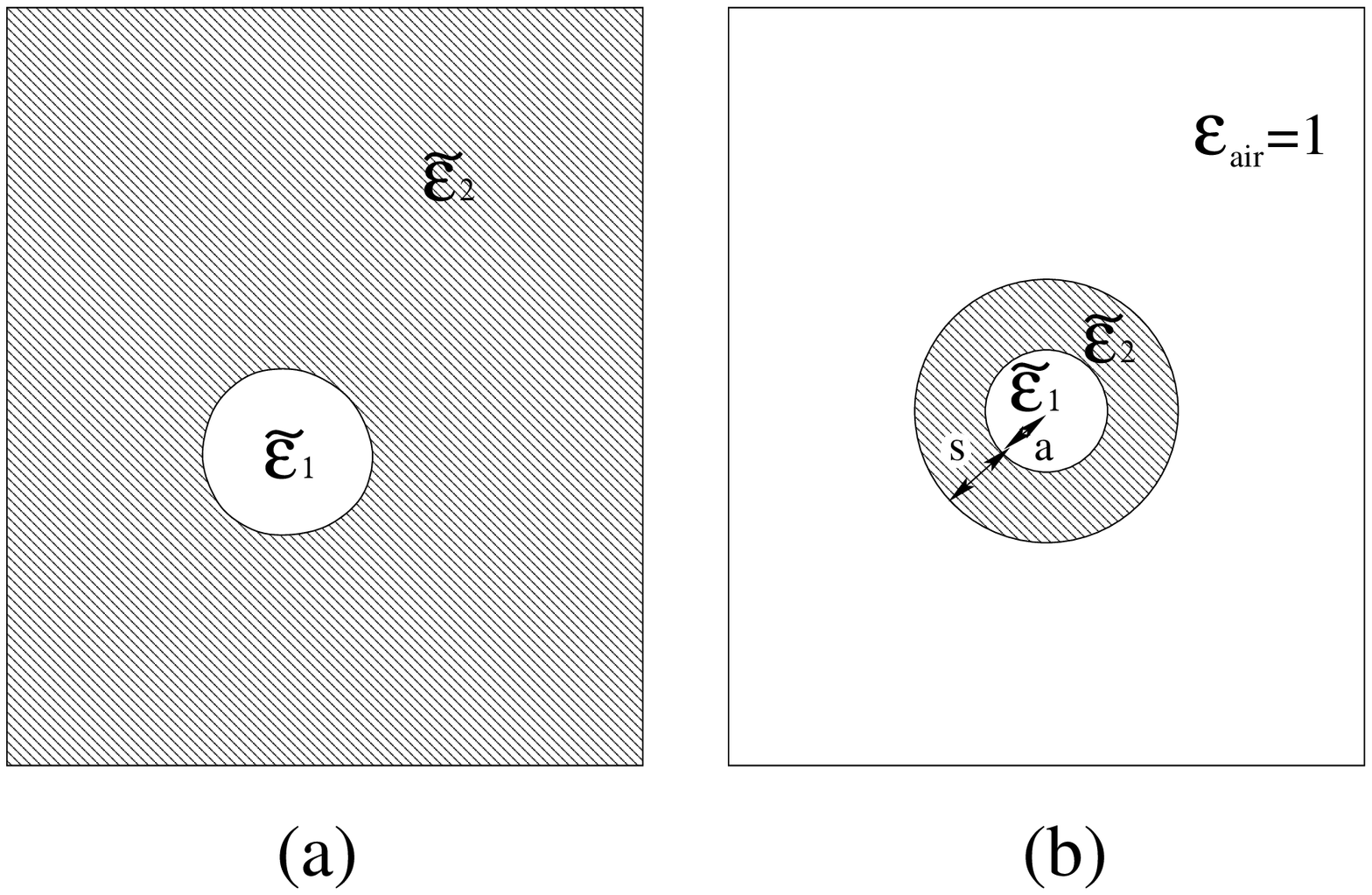}
\end{center}
\caption{}
\end{figure}
\newpage
\begin{figure}
\epsfxsize=15cm \epsfysize=9cm \epsfbox{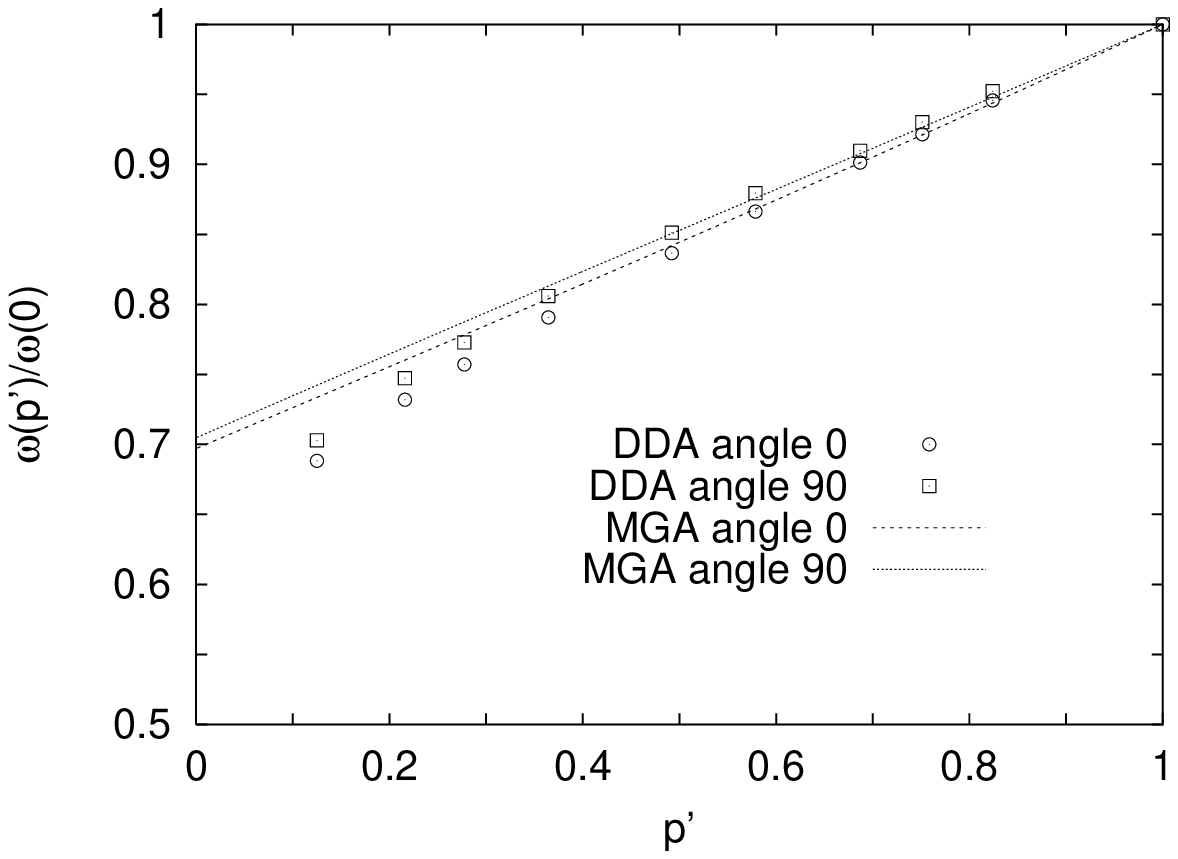}
\caption{}
\end{figure}

\end{document}